\documentclass[pre,amsmath,onecolumn, showpacs, superscriptaddress,10pt]{revtex4-1}
\usepackage{graphicx}
\usepackage{amsfonts}
\usepackage{color}

\newcommand{\bea}{\begin{eqnarray}}
\newcommand{\eea}{\end{eqnarray}}
\newcommand{\be}{\begin{equation}}
\newcommand{\ee}{\end{equation}}

\begin{document}

\title{Number of distinct and common sites visited by $N$ independent random walkers}

\author{Satya N. Majumdar}
\affiliation{Laboratoire de Physique Th\'eorique et Mod\`eles Statistiques (UMR 8626 du CNRS),
Universit\'e Paris-Sud, B\^at.\ 100, 91405 Orsay Cedex, France}
\author{Gr\'egory Schehr}
\affiliation{Sorbonne Universit\'e, Laboratoire de Physique Th\'eorique et Hautes Energies, CNRS UMR 7589, 4 Place Jussieu, 75252 Paris Cedex 05, France}

\begin{abstract}
In this Chapter, we consider a model of $N$ independent random walkers, each of duration $t$, and each starting
from the origin, on a lattice in $d$ dimensions. We focus on two observables, namely $D_N(t)$ and $C_N(t)$
denoting respectively the number of distinct and common sites visited by the walkers. For large $t$, where the 
lattice random walkers converge to independent Brownian motions, we compute exactly the mean $\langle D_N(t) \rangle$ 
and $\langle C_N(t) \rangle$. Our main interest is on the $N$-dependence of these quantities. While for $\langle D_N(t) \rangle$ 
the $N$-dependence only appears in the prefactor of the power-law growth with time, a more interesting behavior emerges for $\langle C_N(t) \rangle$. For this latter case, we show that there is a ``phase transition'' in the $(N, d)$ plane where the two critical line $d=2$ and $d=d_c(N) = 2N/(N-1)$ separate three
phases of the growth of $\langle C_N(t)\rangle$. The results are extended to the mean number of sites visited exactly by $K$ of the $N$ walkers. Furthermore in $d=1$, the full distribution of $D_N(t)$ and $C_N(t)$ are computed, exploiting a mapping to the extreme value statistics. Extensions to two other models, namely $N$ independent Brownian bridges and $N$ independent resetting Brownian motions/bridges are also discussed.
\end{abstract}

\vspace{0.5cm}

\maketitle

\section{Introduction}

In elementary set theory, two fundamental concepts are
the {\em union} and the {\em intersection} of a number of $N$ sets.
While the union consists of all {\em distinct} elements of the
collection of sets, the intersection consists of {\em common} elements
of all the sets. These two notions appear naturally in
everyday life: for example the area of common knowledge or the whole range of
different interests amongst the members of
a society would define respectively its stability and activity.
In an habitat of $N$ animals, the union of the territories
covered by different animals sets the geographical range of the habitat,
while the intersection refers to the common area ({\it e. g.} a water body)
frequented by all animals. In statistical physics, these two objects are modeled respectively by the
number of distinct and common sites visited by $N$ random walkers (RWs) on a $d$-dimensional hyper cubic lattice.
The knowledge about the number of distinct sites  has applications ranging
from the annealing of defects in crystals \cite{BD,Beeler} and
relaxation processes \cite{Blu,Cze,Bor,Cond} to the spread of populations
in ecology \cite{E-K, Pie} or to the dynamics of web annotation systems~\cite{Cattuto}. 

  
Dvoretzky and Erd\"os \cite{DE} first studied the average number 
of distinct sites $\langle S_1(t) \rangle$ visited 
by a single $t$-step RW in $d$-dimensions, subsequently
studied in \cite{Vineyard,MW,FVW}. Larralde {\it et al.} generalized this to $N$ independent 
$t$-step walkers moving on a $d$-dimensional lattice \cite{Larralde}. They found three regimes of growth (early, intermediate and late) for 
the average number of distinct sites
$\langle S_N(t) \rangle$ as a function of time. These three regimes are separated
by two $N$-dependent times scales~\cite{Larralde}. In particular they showed that in $d=1$ and $t\gg \sqrt{\log N}$, 
$\langle S_N(t) \rangle \propto \sqrt{4D~t~\log N}$ where 
$D$ is the diffusion constant of a single walker. Recently Majumdar and Tamm \cite{Majtam} studied 
the complementary quantity, the number of common sites $W_N(t)$ visited by $N$ walkers, each of $t$ steps, and
found analytically a 
rich asymptotic late time growth of the average $\langle W_N(t)\rangle$. In the $(N-d)$ plane they found
three distinct phases separated by two critical lines $d=2$ and $d_c(N)=2N/(N-1)$, with $\langle W_N(t)\rangle\sim t^{\nu}$
at late times where $\nu=d/2$ (for $d<2$), $\nu=N-d(N-1)/2$ [for $2<d<d_c(N)$] and
$\nu=0$ [for $d>d_c(N)$]~(see also~\cite{turban}). In particular,
in $d=1$, $\langle W_N(t)\rangle \sim \sqrt{4Dt}$ with a $N$-dependent prefactor.
While the mean number of distinct and common sites visited by $N$ independent RW's is now well studied in all dimensions,
computing their full distribution is highly nontrivial. Only recently the full distributions of both $D_N(t)$ and $C_N(t)$ were
computed exactly in $1d$ and an interesting link to extreme value statistics was established \cite{KMS2013}.  

In this book chapter, we will first provide a comprehensive and pedagogical introduction to computing the mean number of distinct
and common sites $\langle D_N(t) \rangle$ and $\langle C_N(t) \rangle$ visited by $N$ independent RWs in all dimensions. We will show that
both $\langle D_N(t) \rangle$ and $\langle C_N(t) \rangle$ can be expressed in terms of a central quantity  $p(\vec{x},t)$ denoting the probability
that a site $\vec{x}$ is visited by a single $t$-step walker, that starts at the origin. We will then analyse the scaling behavior of $p(\vec{x},t)$ in
all dimensions and use this to compute the asymptotic large $t$ behavior of $\langle D_N(t) \rangle$ and $\langle C_N(t) \rangle$. Next we will focus on
$d=1$ and demonstrate how to compute the full distribution of $D_N(t)$ and $C_N(t)$ and also establish the interesting link to extreme value statistics.  
In particular, we will show that, for large $N$, the scaling form of the distributions of $\langle D_N(t) \rangle$ and $\langle C_N(t) \rangle$ can be 
expressed in terms of two well known functions (namely Gumbel and Weibull) that appear in extreme value theory. Some more recent extensions of these
techniques will also be discussed at the end. The results that we discuss in this book chapter have already been published elsewhere but, here, we gather
all the results together with unifying notations and methods. We also take this opportunity to provide some more recent applications and
some perspective for future works.

\section{The mean number of distinct and common sites in $d$ dimensions}\label{sec:method}

\begin{figure}[t]
\centering
\includegraphics[width = 0.7\linewidth]{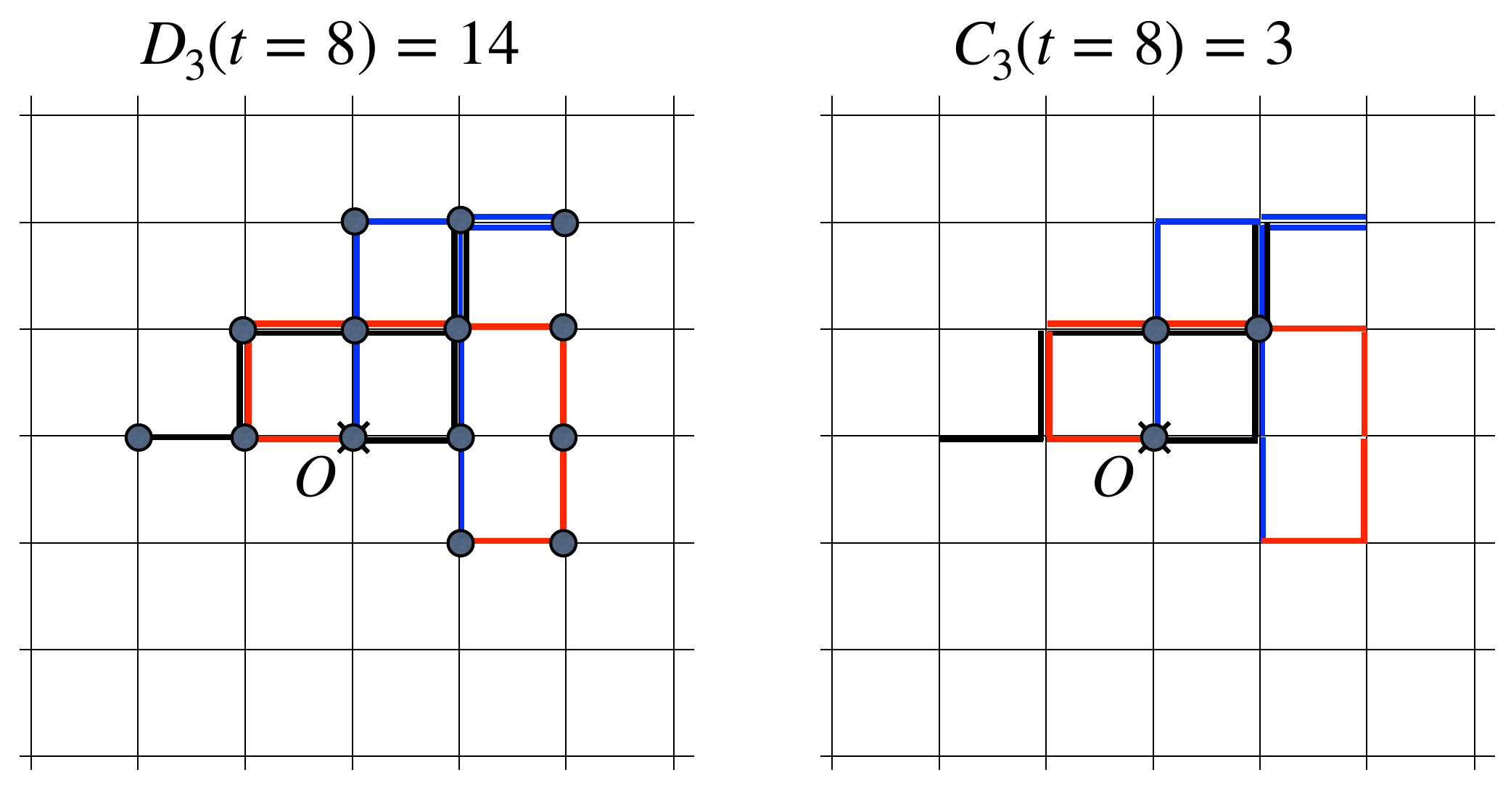}
\caption{Sketch of one realization of $N=3$ lattice random walks starting from the origin, each being of duration $t=8$ steps. On the panel we have marked with dots the distinct sites visited by the $N=3$ random walkers, their number being $D_3(t=8) = 14$. On the right panel we have marked instead the number of common sites visited by the $N=3$ walkers, their number being $C_3(t=8) = 3$.}\label{Fig_intro}
\end{figure}

We first consider a single random walker on a $d$-dimensional hyper cubic lattice. The walker starts
at the origin. At each discrete time-step, the walker hops from a given site to anyone of the $2d$ neighbours
with equal probability $1/(2d)$. Next we consider $N$ such walkers, all starting at the origin and evolving 
independently up to $t$ steps (see Fig. \ref{Fig_intro}). We denote by $D_N(t)$ as the number of sites that are
visited by at least one walker and by $C_N(t)$ the number of sites that have been visited by {\it all} walkers. Both $D_N(t)$ and $C_N(t)$
are random variables and fluctuate from sample to sample. Our main 
interest is to compute the statistics of $D_N(t)$ and $C_N(t)$. Note that for $N=1$, we have $D_1(t) = C_1(t)$, but for $N>1$, they are quite different. 
In this section, we will focus on the mean values $\langle D_N(t) \rangle$
 and $\langle C_N(t) \rangle$. 
 
 We now show how both $\langle D_N(t) \rangle$ and $\langle C_N(t) \rangle$ can be expressed in terms of a central single walker quantity $p(\vec{x},t)$
 denoting the probability that the site $\vec{x}$ is visited by one walker up to time $t$. In order to establish this connection, 
we consider a single walker starting at the origin and, following Ref. \cite{Majtam}, we introduce a binary variable
 \bea \label{def_bin}
 \sigma(\vec{x},t) = 
 \begin{cases}
 &1 \quad, \quad {\rm if}\; \vec{x} \;  {\rm has \; been \; visited \; by \; the \; walker \; up \; to \; time\;} t \;, \\
 &0 \quad, \quad {\rm otherwise} \;.
 \end{cases}
 \eea
Then for a single walker, the random variable $D_1(t) = C_1(t)$ can be written as
\bea \label{D1}
D_1(t) = C_1(t) = \sum_{\vec{x}}  \sigma(\vec{x},t) \;.
\eea
Taking the average on both sides, we get
\bea \label{av_D1}
\langle D_1(t) \rangle =\langle C_1(t) \rangle = \sum_{\vec{x}}  p(\vec{x},t) \;,
\eea
where $p(\vec{x},t) = \langle \sigma(\vec{x},t)\rangle$ is the probability that the site $\vec{x}$ is visited by the walker
within time $t$. Since the walker starts at the origin $\vec{x} = \vec{0}$ 
\bea \label{p0t}
p(\vec{0},t) = 1 \;.
\eea

We now consider $N$ such independent walkers. Using the fact that they all start at the origin, and the fact that they are independent, 
the probability that a site $\vec{x}$ is {\it not} visited by any one of the $t$-step walkers is simply $(1-p(\vec{x},t))^N$. Hence, the probability that
the site $\vec{x}$ is visited by at least one of the walkers is $1-(1-p(\vec{x},t))^N$. Summing over all the sites we get
\bea \label{av_DN}
\langle D_N(t) \rangle = \sum_{\vec{x}} \left[1-(1-p(\vec{x},t))^N\right] \;.
\eea 
Similarly, the probability that a site $\vec{x}$ is visited by all the $t$-step walkers is $(p(\vec{x},t))^N$. Consequently, summing over $\vec{x}$, we get
\bea \label{av_CN}
\langle C_N(t) \rangle = \sum_{\vec{x}}  (p(\vec{x},t))^N \;.
\eea
Thus to compute $\langle D_N(t) \rangle$ and $\langle C_N(t) \rangle$, we need to compute one single central quantity, namely $p(\vec{x},t)$. Even though 
this quantity $p(\vec{x},t)$ can be computed for a random walker on a lattice evolving in discrete time, it turns out that the large time behaviors of 
$\langle D_N(t) \rangle$ and $\langle C_N(t) \rangle$ only require the asymptotic behaviour of $p(\vec{x},t)$ for large $x=|\vec{x}|$ and large $t$. This large distance and late time behaviors of $p(\vec{x},t)$ can be derived directly by replacing the lattice random walker in discrete time by a Brownian motion in 
continuous space and of duration $t$ in continuous time. Consequently, the discrete sums over $\vec{x}$ in Eqs. (\ref{av_DN}) and (\ref{av_CN}) can be replaced by integrals over continuous space. Thus, for both quantities $\langle D_N(t) \rangle$ and $\langle C_N(t) \rangle$, the central object is to compute $p(\vec{x},t)$ for a single Brownian motion of duration $t$, which we derive in the next subsection. 

In fact, once we know the central quantity $p(\vec{x},t)$, there are other interesting observables, in addition to $\langle D_N(t) \rangle$ and $\langle C_N(t) \rangle$, that can be computed. As an example, we consider $\langle V_{K,N}(t) \rangle$ denoting the
mean number of sites that are visited exactly by $K$ walkers ($1\leq K \leq N$) up to time $t$. A site $\vec{x}$ is visited before $t$ by a single walker with probability $p(\vec{x},t)$. Using the independence of the walkers and summing over all $\vec{x}$, it then follows that
\bea \label{VK}
\langle V_{K,N}(t)\rangle = {N \choose K}\sum_{\vec{x}} \left[p(\vec{x},t)\right]^K [1 - p(\vec{x},t)]^{N-K} \;.
\eea 
For $K=N$, we just have $\langle V_{N,N}(t) \rangle= \langle C_N(t) \rangle$. Thus the mean number of common sites is just a special case of $\langle V_{K,N}(t)\rangle$ with $K=N$.

\subsection{The central quantity $p(\vec{x},t)$} 

Consider then a single Brownian motion of length t and diffusion constant $D$ in $d$-dimensions starting at the origin. We are interested in $p(\vec{x},t)$, the probability that the site $\vec{x}$ is visited (at least once) by the walker up to time $t$. Let $\tau$ denote the last time before $t$ that the site $\vec{x}$ was visited by the walker. Then, using the Markov property of the walk, it follows that
\bea \label{Markov}
p(\vec{x},t) = \int_0^t G(\vec{x},\tau) q(t-\tau)\, d\tau \;,
\eea
where $G(\vec{x},\tau)$ is the propagator of the Brownian motion
\bea \label{propag}
G(\vec{x},\tau) = \frac{1}{(4 \pi D \, \tau)^{d/2}} \, e^{-x^2/(4 D \tau)} \;,
\eea 
measuring the probability density of reaching $\vec{x}$ at time $\tau$, starting from the origin at time $t=0$. The quantity $q(\tau)$ in Eq. (\ref{Markov}) denotes
the probability that the walker, starting at $\vec{x}$, does not to return to $\vec{x}$ up to time $\tau$. Due to the translation invariance of the walk, $q(\tau)$ does not depend on $\vec{x}$ is thus also the probability that the walker, starting at the origin, does not return to the origin up to time $t$.  

The no-return probability $q(\tau)$ for a $d$-dimensional Brownian motion can be computed as follows. It is useful first to relate it to the first-return probability 
$F(\tau)$ to the origin by the relation $q(\tau) = \int_{\tau}^\infty F(\tau')\, d\tau'$. In other words 
\bea \label{dq}
\frac{dq}{d\tau} = - F(\tau) \;.
\eea
We also define their Laplace transforms
\bea \label{LT}
\tilde q(s) = \int_0^\infty q(t)\, e^{-st} \, dt \quad, \quad  \tilde F(s) = \int_0^\infty F(t)\, e^{-st} \, dt \;.
\eea
Taking Laplace transform of Eq. (\ref{dq}) and using $q(0) = 1$, we get the well known relation \cite{Rednerbook,us_book}
\bea \label{LT_2}
\tilde q(s) = \frac{1- \tilde F(s)}{s} \;.
\eea
The Laplace transform of the first-passage probability $\tilde F(s)$ can further be related to the Laplace transform of the propagator $G(\vec{0},t)$ as
follows
\bea \label{rel_F_G}
G(\vec{0},t) = \delta(t) + \sum_{n=1}^\infty \int \left(\prod_{i=1}^n F(t_i) dt_i \right)  \delta\left(\sum_{i=1}^n t_i - t\right) \;,
\eea  
where the $m$-th term ($m \geq 1$) of the sum corresponds to trajectories with exactly $m-1$ returns to the origin before time $t$ and the $m$-th return 
exactly at $t$. The first term $\delta(t)$ just reflects the initial condition that the walker starts at the origin at time $t=0$. Taking Laplace transform on both sides and evaluating the sum as a geometric series gives
\bea \label{LT_rel_F_G}
\tilde G(\vec{0},s) = \frac{1}{1 - \tilde F(s)} \;.
\eea
Eliminating $1 - \tilde F(s)$ between Eqs. (\ref{LT_2}) and (\ref{LT_rel_F_G}) gives
\bea\label{LT_3}
\tilde q(s) = \frac{1}{s \, \tilde G(\vec{0},s)} \;.
\eea
Let us recall that $\tilde G(\vec{0},s) = \int_0^\infty e^{-s\,t}  G(\vec{0},t) \, dt$ where the propagator $G(\vec{x},t)$ is given in (\ref{propag}). If we substitute $\vec{x} = \vec{0}$ in Eq. (\ref{propag}), one would get $G(\vec{0},t) = 1/(4 \pi D\, t)^{d/2}$. However, as we will see later, one would need a lattice cut-off $\vec{a}$ in order to regularise the integrals over $t$, in particular for $d>2$. Hence, we will use the following regularised expression for $G(\vec{0},t)$, 
\bea \label{reg}
G(\vec{0},t) \simeq \frac{1}{(4 \pi D t)^{d/2}} \, e^{-\frac{a^2}{4 D t}} \;.
\eea
We will see later that, for $d<2$, we can take eventually the $a \to 0$ limit and the answer will be finite. In contrast, for $d>2$, we need to keep a nonzero cut-off
$a$ and the result for $p(\vec{x},t)$ will depend explicitly on the cut-off $a$.

\subsection{The scaling analysis of $p(\vec{x},t)$ in all dimensions}

From Eq. (\ref{Markov}), we see that, to extract the scaling behavior of $p(\vec{x},t)$ for large $|\vec{x}|$ and large $t$, we need to know the behavior of the no-return probability to the origin up to time $t$, i.e., $q(t)$ for large $t$. Below, we first extract the large $t$ behavior of $q(t)$ in all dimensions and then substitute this asymptotic behavior in Eq. (\ref{Markov}) to extract the scaling behavior of $p(\vec{x},t)$. This is the approach that was used in Ref. \cite{Majtam}.

\subsubsection{Asymptotic behavior of $q(t)$ for large $t$}

For this asymptotic analysis of $q(t)$, we start from its exact Laplace transform in Eq. (\ref{LT_3}), with $\tilde G(\vec{0},s)$ given by
\be \label{tildeG_expl}
\tilde G(\vec{0},s) \simeq \int_0^\infty e^{-s t} \,  \frac{1}{(4 \pi D t)^{d/2}} \, e^{-\frac{a^2}{4 D t}}\, dt  = \frac{1}{(4 \pi D)^{d/2}} \frac{1}{s^{1-d/2}} \, \int_0^\infty \frac{e^{-\frac{a^2\,s}{4 D y}}\,e^{-y}}{y^{d/2}}\, dy \;,
\ee
where we used Eq. (\ref{reg}) in the first relation and then made a change of variable $y = s\,t$ in the second equality. We would 
now like to consider the small $s$ behavior of the integral in Eq. (\ref{tildeG_expl}) in 
three separate cases: (i) $d<2$, (ii) $d>2$ and (iii) $d=2$. 

\begin{itemize}

\item[(i)]{$d<2$: In this case, we can take the limit $s \to 0$ in the integral, which remains convergent. 
This then gives, to leading order as $s \to 0$ 
\bea \label{dless2}
\tilde G(\vec{0},s) \simeq \frac{\Gamma\left( 1 - \frac{d}{2}\right)}{(4 \pi D)^{d/2} s^{1-d/2}} \;.
\eea
Substituting this in Eq. (\ref{LT_3}) gives, for small $s$,
\bea \label{LT_4}
\tilde q(s) \sim \frac{(4 \pi D)^{d/2}}{\Gamma \left(1 - d/2 \right)} \frac{1}{s^{d/2}} \;.
\eea
To invert this Laplace transform, we use the identity
\bea \label{id_LT}
{\cal L}^{-1}_{s \to t} \left[ \frac{1}{s^\alpha}\right] = \frac{1}{\Gamma(\alpha)\, t^{1-\alpha}}  \quad, \quad {\rm for} \quad \alpha > 0\;.
\eea
Using this identity with $\alpha = d/2$ gives the 
leading large $t$ behavior of $q(t)$ as
\bea \label{qt_dless2}
q(t) \simeq \frac{A_d}{t^{1-d/2}} \quad, \quad A_d = \frac{(4 \pi D)^{d/2}}{\pi}\, \sin{\left( \frac{\pi d}{2}\right)} \;.
\eea
Note that the leading large $t$ behavior of $q(t)$ is independent of the cut-off $a$ for $d<2$. 
}

\item[(ii)]{\hspace*{0.cm} $d>2$: In this case, we see from the definition of $\tilde G(\vec{0},s)$ in Eq. (\ref{tildeG_expl}) that, when $s \to 0$, we need a finite cut-off $a>0$ for the integral to be convergent. Thus setting $a=0$ gives, for small $s$,
\bea \label{tildeG_expldge2}
\tilde G(\vec{0},s)  \underset{s \to 0}{\longrightarrow} \int_0^\infty \frac{1}{(4 \pi D t)^{d/2}} \, e^{-\frac{a^2}{4 D t}}\, dt \equiv \frac{1}{E_d} \;,
\eea
where the constant $E_d$ depends explicitly on the cut-off $a$. Substituting this in Eq. (\ref{LT_3}) gives, for small $s$,
\bea \label{LT_5}
\tilde q(s) \simeq \frac{E_d}{s} \;.
\eea
Inverting trivially the Laplace transform gives, for large $t$
\bea \label{qt_ge2}
q(t) \simeq E_d \quad, \quad {\rm as} \quad t \to \infty \;.
\eea
Thus $E_d$ is exactly the no-return or eventual ``escape'' probability of the Brownian walker in $d>2$

}

\item[(iii)]{\hspace*{0.cm} $d=2$: In this case, we start from the definition 
\bea \label{Gtilde_d2}
\tilde G(\vec{0},s) \simeq \int_0^\infty \frac{1}{4 \pi D t}\, e^{-\frac{a^2}{4 Dt}} \,e^{-st} \, dt \;.
\eea
In this case, if we directly set $s=0$, we see that we need a nonzero cut-off $a>0$ in order that the integral is convergent. However, the leading small $s$ behavior turns out to be independent of the cut-off $a$ as we show now. Indeed, taking derivative of Eq. (\ref{Gtilde_d2}) with respect to $s$ gives
\bea \label{Gtilde_d2.2}
\frac{d G(\vec{0},s)}{ds} = - \frac{1}{4\pi D}\int_0^\infty e^{-\frac{a^2}{4 Dt}}\, e^{-s t} dt = - \frac{1}{4\pi D\,s} \int_0^\infty e^{- \frac{a^2 s}{4D}} \,e^{-y} dy 
\eea
We can now take the limit $s \to 0$ limit inside the integral since it remains convergent and this gives $\frac{d G(\vec{0},s)}{ds} \simeq - \frac{1}{4 \pi D} \frac{1}{s}$. Integrating it back, we get, to leading order as $s \to 0$, 
\bea \label{Gtilde_d2.3}
G(\vec{0},s) \simeq - \frac{1}{4 \pi D}\, \ln s \;, 
\eea
which is clearly independent of the cut-off $a$. We substitute this back in Eq. (\ref{LT_3}) to obtain the leading small $s$ behavior 
\bea \label{LT_6}
\tilde q(s) \simeq - \frac{4\pi D}{s\, \ln s} \;.
\eea
Inverting this Laplace transform, we get for large $t$
\bea \label{qt_deq2}
q(t) \simeq \frac{4 \pi D}{\ln t} \;.
\eea
}
\end{itemize}

Thus, to summarise, the leading large $t$ behavior of $q(t)$ in different dimensions is given by
\bea \label{summary_qt}
q(t) \simeq 
\begin{cases}
&\dfrac{A_d}{t^{1-d/2}} \quad, \quad d<2\;, \\
& \\
&\dfrac{4 \pi D}{\ln t} \quad, \quad d = 2 \;,\\
& \\
& E_d \quad, \quad \quad d>2 \;,
\end{cases}
\eea  
where the constants $A_d$ and $E_d$ are given respectively in Eqs. (\ref{qt_dless2}) and (\ref{tildeG_expldge2}). This shows that, for $d \leq 2$, the no-return probability $q(t) \to 0$ as $t \to \infty$, indicating that the walk is recurrent. In contrast, for $d>2$, it approaches a nonzero constant, indicating that the walk can escape to infinity with a finite probability $E_d$. Thus this result in Eq. (\ref{summary_qt}) illustrates the well known \cite{Rednerbook,Feller} fact that the Brownian walker is recurrent for $d \leq 2$, while it is transient for~$d>2$.

\subsubsection{Asymptotic behavior of $p({\vec x},t)$ for large $|{\vec x}|$ and large $t$}

We start with Eq. (\ref{Markov}) and first make a change of variable $\tau = t u$, with $0<u<1$. This gives
\bea \label{Markov.2}
p(\vec{x},t) = \frac{t^{1-d/2}}{(4 \pi D)^{d/2}} \int_0^1 \frac{e^{-z^2/u}}{u^{d/2}}\, q\left( t(1-u)\right) \, du\quad, \quad {\rm where} \quad \; z = \frac{|\vec{x}|}{\sqrt{4 D t}}.
\eea
Thus, keeping the scaling variable $z$ fixed, if we take the large $t$ limit, the behaviour of the integral is controlled by the large argument behavior of the no-return probability $q(\tau)$. Hence we can directly substitute in the integral the large argument behaviour of $q(\tau)$ from Eq.~(\ref{summary_qt}). Thus, once again, we analyse three different cases: (i) $d<2$, (ii) $d>2$ and (iii) $d=2$.

\begin{figure}[t]
\includegraphics[width = \linewidth]{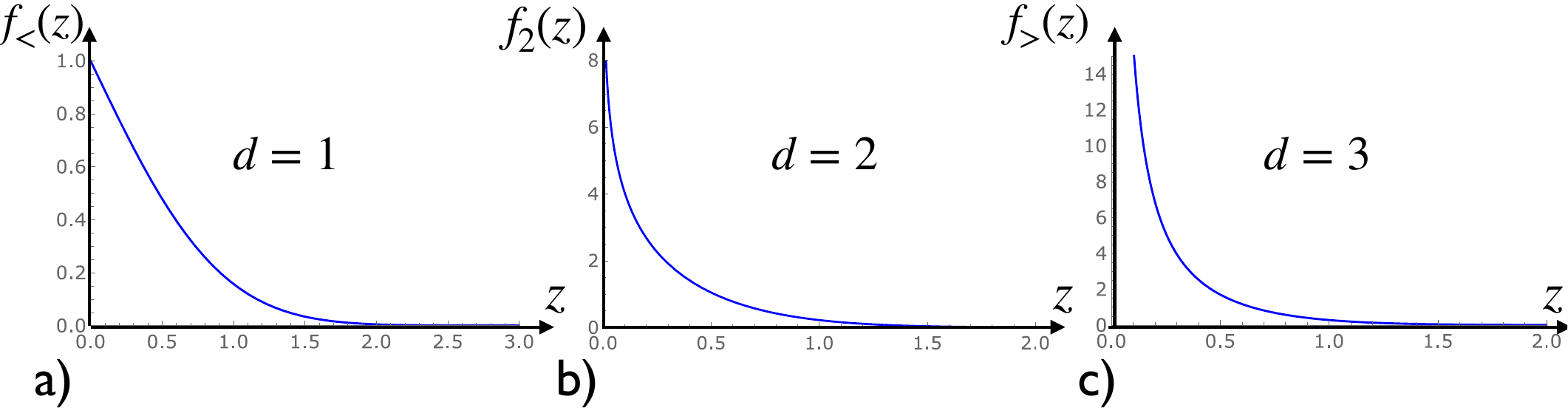}
\caption{{\bf a):} Plot of the scaling function $f_<(z)$ vs $z$ for $d=1$. {\bf b)} Plot of the scaling function $f_2(z)$ vs $z$ ($d=2$). {\bf c):} Plot of the scaling function $f_>(z)$ vs $z$ for $d=3$.} \label{Fig_f}
\end{figure}

\begin{itemize}
\item[(i)]{$d<2:$ In this case substituting the behavior of $q(\tau)$ from the first line of Eq. (\ref{summary_qt}) in Eq. (\ref{Markov.2}) we get
\bea \label{p_dless2}
&&p(\vec{x},t) \simeq f_{<}\left(z = \frac{|\vec{x}|}{\sqrt{4 D t}} \right) \;, \nonumber \\
&&{\rm where} \quad f_{<}(z) = \frac{\sin{\left( \frac{\pi d}{2}\right)}}{\pi} \int_0^1 e^{-z^2/u} u^{-d/2} (1-u)^{d/2-1} \, du \;.
\eea
It is easy to show that the function $f_{<}(z)$ decreases monotonically with increasing $z$. In Fig. \ref{Fig_f} a) we show a plot of this function for $d=1$ where $f_<(z) = {\rm erfc}(z) = (2/\sqrt{\pi}) \int_z^{\infty} e^{-u^2} du$. For general $d<2$, the scaling function has the following asymptotic behaviors
\bea \label{asympt_f<}
f_{<}(z) \simeq
\begin{cases}
&1 - \dfrac{1}{\Gamma\left(2 - \frac{d}{2}\right)} z^{2-d} \quad, \quad \hspace*{0.5cm} z\to 0\;, \\
&\\
& \dfrac{\sin\left(\frac{\pi d}{2}\right)}{\pi} \dfrac{e^{-z^2}}{z^d} \quad , \hspace*{1.2cm}\quad z \to \infty \;.
\end{cases}
\eea

}

\item[(ii)]{\hspace*{0.cm}$d>2:$ Using $q(t) \simeq E_d$ for large $t$ from Eq. (\ref{summary_qt}) in Eq. (\ref{Markov.2}), we get the following scaling behavior
\bea \label{p_dge2}
&&p(\vec{x},t) \simeq \frac{E_d}{(4 \pi D)^{d/2}}\, t^{1-d/2}\, f_>\left( z =  \frac{|\vec{x}|}{\sqrt{4 D t}} \right) \quad, \nonumber \\
&&{\rm where} \quad f_{>}(z) = \int_0^1 \frac{e^{-z^2/u}}{u^{d/2}}\, du \;.
\eea
A plot of this function for $d=3$ is shown in Fig. \ref{Fig_f} c). The asymptotic behaviors of this scaling function $f_>(z)$ are given by
\bea \label{asympt_f>}
f_{>}(z) \simeq
\begin{cases}
&\dfrac{\Gamma\left(d/2-1\right)}{z^{d-2}} \quad, \quad z \to 0\;, \\
&\\
& \dfrac{e^{-z^2}}{z^2} \quad , \quad \hspace*{0.9cm} z \to \infty \;.
\end{cases}
\eea

}

\item[(iii)]{\hspace*{0.cm}$d=2:$ In this case, substituting $q(t) \simeq (4\pi D)/\ln t$ for large $t$ in Eq. (\ref{Markov.2}) gives the scaling behavior
\bea \label{p_eq2}
p(\vec{x},t) \simeq \frac{1}{\ln t} f_2 \left( z =  \frac{|\vec{x}|}{\sqrt{4 D t}} \right) \quad, \quad {\rm where} \quad f_{2}(z) = \int_{0}^1 \frac{e^{-z^2/u}}{u}\, du \;.
\eea
A plot of this function is shown in Fig. \ref{Fig_f} b). The scaling function $f_2(z)$ has the following asymptotic behaviors
\bea \label{asympt_f2}
f_{2}(z) \simeq
\begin{cases}
&-2 \ln z \quad, \quad z \to 0\;, \\
&\\
& \dfrac{e^{-z^2}}{z^2} \quad , \quad \hspace*{0.25cm} z \to \infty \;.
\end{cases}
\eea

}

\end{itemize}

Thus, summarising the scaling forms for $p(\vec{x},t)$ in different dimensions, we get \cite{Majtam}
\bea \label{summary_pxt}
p(\vec{x},t) \simeq
\begin{cases}
& f_<\left(\frac{|\vec{x}|}{\sqrt{4 D t}} \right) \quad, \hspace*{2.3cm}\quad d<2 \;,\\
& \\
& \dfrac{1}{\ln t} f_2 \left(\frac{|\vec{x}|}{\sqrt{4 D t}} \right) \quad, \hspace*{1.9cm}\quad d =2 \; \\
& \\
& \frac{E_d}{(4 \pi D)^{d/2}}\, t^{1-d/2}\, f_>\left(\frac{|\vec{x}|}{\sqrt{4 D t}} \right) \quad, \quad d>2 \; \;,
\end{cases}
\eea
where the scaling functions $f_<(z)$, $f_2(z)$ and $f_>(z)$ are given respectively in Eqs. (\ref{p_dless2}), (\ref{asympt_f2}) and (\ref{asympt_f>}).

\subsection{The asymptotic behavior of $\langle D_N(t) \rangle$ and $\langle C_N(t) \rangle$}

In this subsection, we use the scaling behavior of $p(\vec{x},t)$ from Eq. (\ref{summary_pxt}) in Eqs. (\ref{av_DN}) and (\ref{av_CN}) to compute the mean number of distinct and common sites visited by $N$ Brownian walkers up to time $t$ for large $t$. 

\subsubsection{$\langle D_N(t) \rangle$ for large $t$}

The average number of distinct sites visited up to time $t$ clearly depends on the dimension $d$. We thus consider the three cases separately: (i) $d<2$, (ii) $d>2$ and (iii) $d=2$.

\begin{itemize}

\item[(i)]{$d<2$: We substitute the scaling behavior of $p({\vec x},t)$ from the first line of Eq. (\ref{summary_pxt}) in Eq. (\ref{av_DN}). Performing the change of variable $z = |{\bf x}|/\sqrt{4 Dt}$ and using the spherical symmetry, we get
\bea \label{DNless2}
\langle D_N(t) \rangle \simeq B_{N}(d)\, t^{d/2} \quad {\rm as} \quad t \to \infty \;,
\eea
where the prefactor $B_N(d)$ is given by
\bea \label{BN_less2}
B_N(d) = (4 D)^{d/2} S_d \, \int_0^\infty dz\, z^{d-1} \left[1 - \left(1-f_<(z)\right)^N \right] \;.
\eea
Here $S_d = 2 \pi^{d/2}/\Gamma(d/2)$ is the surface area of the unit sphere in $d$ dimensions and the scaling function $f_<(z)$ is given in Eq. (\ref{p_dless2}). It is hard to compute this prefactor $B_N(d)$ in Eq. (\ref{BN_less2}) explicitly for general $N$. However, one can easily extract the asymptotic large $N$ behavior of $B_N(d)$. For large $N$, the integral in Eq. (\ref{BN_less2}) is dominated by the large $z$ behavior of $f_<(z)$ given in Eq. (\ref{asympt_f<}). Substituting this asymptotic behavior in Eq. (\ref{BN_less2}), one finds 
\bea \label{inter_less2}
\left(1-f_<(z)\right)^N \sim \exp{\left[- N  \dfrac{\sin\left(\frac{\pi d}{2}\right)}{\pi} \dfrac{e^{-z^2}}{z^d}\right]} \;.
\eea
As a function of $z$, the right hand side of Eq. (\ref{inter_less2}) approaches $1$ as $z \to \infty$ and vanishes as $z \to 0$. As $N \to \infty$, this  jump from $0$ to $1$ occurs at $z \simeq \sqrt{\ln N}$ where the argument of the exponential vanishes. Thus, for large $N$, we can approximate the right hand side of Eq. (\ref{inter_less2}) by a step function $\theta\left(\sqrt{\ln N}-z\right)$. Using this large $N$ approximation in Eq. (\ref{BN_less2}), we see that we can cut-off the integral at $z = \sqrt{\ln N}$ and this gives, to leading order for large $N$, 
\bea \label{BN_less2.1}
B_N(d) \simeq \frac{(4 \pi D)^{d/2}}{\Gamma \left( \frac{d}{2}+1\right)}\, (\ln N)^{d/2} \quad {\rm as} \quad N \to \infty \;.
\eea
}

\item[(ii)]\hspace*{0.cm}{$d>2:$  Here also, we substitute the scaling behavior of $p({\vec x},t)$ from the third line of Eq. (\ref{summary_pxt}) in Eq. (\ref{av_DN}). This gives, using the spherical symmetry
\be \label{DNge2}
\langle D_N(t) \rangle \simeq S_d\,\int_0^\infty dx\, x^{d^1}\, \left[1 - \left(1- \frac{E_d}{(4 \pi D)^{d/2}}\, t^{1-d/2}\, f_>\left(\frac{|\vec{x}|}{\sqrt{4 D t}} \right)\right)^N \right] \;.
\ee
For $d>2$ and large $t$, the amplitude of the term containing $f_>(z)$ is small and hence, expanding in Taylor series and keeping only the leading term gives
\bea \label{DNge2.2}
\langle D_N(t) \rangle &\simeq&  S_d \frac{N\, E_d}{(4 \pi D)^{d/2}} \int_0^\infty dx \frac{x^{d-1}}{t^{d/2-1}} f_>\left( \frac{x}{\sqrt{4 D t}}\right) \nonumber \\
&=& E_d\,N\,t\, \frac{S_d}{\pi^{d/2}} \int_0^\infty dz \, z^{d-1} f_>(z) \;.
\eea
Performing the integral over $z$, and using $S_d = 2 \pi^{d/2}/\Gamma(d/2)$ gives the very simple answer
\bea \label{DNge2.3}
\langle D_N(t) \rangle \simeq E_d\,N\,t \;.
\eea
There is an alternative way of arriving at the same result. Starting from the definition in Eq. (\ref{av_DN}), we see that, at late times, $p(\vec{x},t) \to 0$ as $t^{1-d/2}$ for $d>2$, for fixed $z = |\vec{x}|/\sqrt{4 D t}$ (see the third line of Eq. (\ref{summary_pxt})). Expanding in Taylor series and keeping only the leading term gives
\bea \label{DNge2.4}
\langle D_N(t) \rangle \simeq N \int d^d x \; p(\vec{x},t) \;. 
\eea
Substituting the expression for $p(\vec{x},t)$ from Eq. (\ref{Markov}) and using $\int d^d x\, G(\vec{x},t) = 1$ gives
\bea \label{DNge2.5}
\langle D_N(t) \rangle \simeq N \int_0^t q(t-\tau) d\tau =  N \int_0^t q(\tau) d\tau \;.
\eea
Using the fact that, for $d>2$, the no-return probability $q(\tau) \to E_d$ for large $\tau$ [see the third line of Eq. (\ref{summary_qt})], immediately reproduces the result in Eq. (\ref{DNge2.3}). Physically, this result has the following implication. Consider first a single walker at late times $t$ in $d>2$ dimensions. The total number of sites visited is $t$, while in this limit the probability that a given site is not revisited approaches $E_d$ at late times. Hence the mean number of distinct sites visited by a single walker approaches $E_d\,t$ asymptotically.  
For large $N$ and $d>2$, the independent walkers hardly overlap and each of them visits on an average $E_d\, t$ distinct sites. This gives the result in Eq. (\ref{DNge2.3}). 
} 

\item[(iii)]{\hspace*{0.cm} $d=2$: In this case, we see from the second line of Eq. (\ref{summary_pxt}), that for fixed $z = |\vec{x}|/\sqrt{4 Dt}$, the probability
$p(\vec{x},t)$ still decays to zero as $t \to \infty$, albeit very slowly as $1/\ln{t}$. Hence, as in the $d>2$ case above, we expand Eq. (\ref{av_DN}) in a Taylor series for small $p(\vec{x},t)$ and keep only the leading term, which again gives, to leading order for large $t$,  
\bea \label{DNge2.6}
\langle D_N(t) \rangle \simeq N \int_0^t q(t-\tau) d\tau =  N \int_0^t q(\tau) d\tau \;.
\eea
Using the asymptoptic behavior of $q(\tau) \simeq 4 \pi D/\ln \tau$ from the second line of Eq. (\ref{summary_qt}), we get, to leading order for large $t$
\bea\label{DNge2.7}
\langle D_N(t) \rangle \simeq  N \frac{4 \pi D \,t}{\ln t} \;.
\eea
Thus the asymptotic non-overlapping of the number of distinct sites visited by $N$ independent walkers remains true even for $d=2$, as in the $d>2$ case.
}

\end{itemize}

Let us finish this subsection by summarising the leading large $t$ behavior of $\langle D_N(t) \rangle$ in different dimensions. We obtain 
\bea \label{DN_summary}
\langle D_N(t) \rangle \simeq
\begin{cases}
& B_N(d)\, t^{d/2} \quad, \quad d<2  \;,\\
& \\
&  N \dfrac{4 \pi D \,t}{\ln t} \quad, \quad \quad d= 2 \;,\\
&\\
& E_d\,N\,t \quad, \quad \quad \;\; d>2 \;,
\end{cases}
\eea
where $B_N(d)$ and $E_d$ are given respectively in Eqs. (\ref{BN_less2}) and (\ref{tildeG_expldge2}). Thus the temporal growth of $\langle D_N(t) \rangle$ for large $t$ for $N$ independent walkers is identical to that of a single walker and the $N$-dependence emerges only in the prefactor of this asymptotic growth. In Ref. \cite{Larralde}, the time dependence of $\langle D_N(t) \rangle$ was analysed in $d=1, 2$ and $3$ using a slightly different approach. The late time results in Eq. (\ref{DN_summary}) are consistent with the results of Ref. \cite{Larralde} for $d=1, 2$ and $3$. In addition, the authors of Ref. \cite{Larralde} also found an intermediate regime where the time-dependence is different from the asymptotic growth. This intermediate regime can also be recovered from our approach, though we do not provide details here.

\subsubsection{$\langle C_N(t) \rangle$ for large $t$}

Here, we derive the asymptotic large $t$ behavior of $\langle C_N(t) \rangle$. As in the case of $\langle D_N(t) \rangle$, the asymptotic behavior of $\langle C_N(t) \rangle$ changes at $d=2$. But as we will see below, that for $d>2$, the asymptotic behavior of $\langle C_N(t) \rangle$ is much richer than that of $\langle D_N(t) \rangle$. In fact, it turns out that there is another critical dimension $d_c(N) = 2N/(N-1)$ such that, for $2<d<d_c(N)$, the mean number of common sites grows as a power law $\langle C_N(t)\rangle \sim t^{\nu}$ for large $t$ where the exponent $\nu = N - d(N-1)/2$ depends on both $d$ and $N$ \cite{Majtam}. In contrast, for $d>d_c(N)$, $\langle C_N(t) \rangle$ approaches a constant at late times. Below we discuss these cases separately.

 \begin{itemize}
 
 \item[(i)]{$d<2$: We start from the formula in Eq. (\ref{av_CN}) and substitute the scaling form of $p(\vec{x},t)$ from the first line in Eq. (\ref{summary_pxt}). Using spherical symmetry, we then get
 \bea \label{CN_dless2}
 \langle C_N(t) \rangle \simeq b_N(d)\, t^{d/2} \quad {\rm as} \quad t \to \infty \;,
 \eea
 where the amplitude $b_N(d)$ is given by
 \bea \label{bN_less2}
 b_N(d) = (4 D)^{d/2}\, S_d \, \int_0^\infty dz\, z^{d-1} \, \left[f_<(z)\right]^N \;.
 \eea
 Once again, it is difficult to compute this integral explicitly for arbitrary $d<2$ and $N$. However, one can extract the large $N$ behavior
 of $b_N(d)$ for fixed $d$ as follows. For large $N$, the integral is dominated by the small $z$ behaviour of $f_<(z)$. To see this, we first substitute the small $z$ behavior of $f_<(z)$ from the first line of Eq. (\ref{asympt_f<}). This gives
 \bea \label{bN_less2.2}
 b_N(d) \simeq (4 D)^{d/2} S_d \int_0^\infty dz\; z^{d-1} \left[1- \frac{1}{\Gamma\left(2-d/2\right)}\, z^{d-2} \right]^N  \;.
 \eea 
 In the limit $N \to \infty$, the leading contribution to the integral comes from the region where $z \sim N^{1/(2-d)}$. In this regime, to leading order for large $N$, one can make the approximation 
\bea \label{approx_dless2}
\left[1- \frac{1}{\Gamma\left(2-d/2\right)}\, z^{d-2} \right]^N \simeq \exp{\left[-\frac{N}{\Gamma\left(2-d/2\right)}\, z^{d-2}\right]} \;.
\eea
Substituting this behaviour back into Eq. (\ref{bN_less2.2}) and performing the integral over $z$ explicitly, we get the leading large $N$ behavior of the amplitude $b_N(d)$ as
\be  \label{bN_less2.3}
b_N(d) \simeq \frac{\tilde b(d)}{N^{\frac{d}{(2-d)}}} \quad {\rm with} \quad \tilde b(d) = (4 D)^{d/2} \, \frac{S_d}{d} \, \Gamma\left( \frac{2}{2-d}\right) \,\left[ \Gamma\left(\frac{4-d}{2}\right)\right]^{\frac{d}{2-d}}
\ee
where we recall that $S_d = 2 \pi^{d/2}/\Gamma(d/2)$ is the surface area of the unit sphere in $d$ dimensions. By comparing with Eq. (\ref{BN_less2.1}), we see that, while both $\langle D_N(t) \rangle$ and $\langle C_N(t) \rangle$ grow as $t^{d/2}$ for large $t$, the amplitude of this growth has very different dependence on $N$. In the case of $\langle D_N(t) \rangle$, the amplitude $B_N(d)$ grows as $(\ln N)^{d/2}$ for large $N$, while for $\langle C_N(t)\rangle$, the amplitude $b_N(d)$ decreases as a power law $\sim N^{-d/(2-d)}$ for large $N$. 
 
 }

\item[(ii)]{\hspace*{0.cm}$d>2$: We now consider the case $d>2$ where the scaling behavior of $p(\vec{x},t)$ is given in the third line of Eq. (\ref{summary_pxt}). We substitute this scaling form in Eq. (\ref{av_CN}), replace the sum over $\vec{x}$ by a continuous integral and perform this integral using the spherical symmetry. This gives
\bea \label{CN_dge2}
\langle C_N(t) \rangle \simeq t^{\nu}\,\frac{E_d^N}{(4 \pi D)^{Nd/2}} {S_d\, (4 D)^{d/2}} \int_0^\infty dz \,z^{d-1}\, [f_>(z)]^N  \;,
\eea
where the exponent $\nu$ is given by 
\bea \label{nu}
\nu = N- \frac{d(N-1)}{2} \;,
\eea
while the scaling function $f_>(z)$ is given in Eq. (\ref{p_dge2}). This result, of course, makes sense only when the integral on the right hand side is convergent. To find the condition of this convergence, we only examine the small $z$ behavior of the integrand [for large $z$, the integral is convergent anyway given the asymptotic behavior given in the second line of Eq. (\ref{asympt_f>})]. As $z \to 0$, from the first line of Eq. (\ref{asympt_f>}), we have $f_>(z) \sim 1/z^{d-2}$. Consequently, the integrand $z^{d-1} \, [f_>(z)]^N \sim z^{2N - d(N-1)-1}$. Thus the integral is convergent in the lower limit $z \to 0$ provides $d < d_c(N)$ where
\bea \label{dcN}
d_c(N) =  \frac{2N}{N-1} \;.
\eea 
Thus, for $d<d_c(N)$, Eq. (\ref{CN_dge2}) predicts that $\langle C_N(t) \rangle$ grows algebraically for large time as
\bea \label{CN_dge2.2}
\langle C_N(t) \rangle \simeq  \alpha_N(d)\, t^{\nu} \;,
\eea
with $\nu = (N-1)(d_c-d)/2 >0$ given in Eq. (\ref{nu}). The amplitude $\alpha_N(d)$ is then given by
\bea \label{alphaN_d}
\alpha_N(d)= S_d\, (4D)^{d/2}\, \frac{E_d^N}{(4\pi D)^{dN/2}}\, \int_0^{\infty} dz\, z^{d-1}\, \left[f_{<}(z)\right]^N \;.
\eea

A different situation occurs for $d>d_c(N)$. First, we note that the scaling form in Eq. (\ref{summary_pxt}) holds for $|{\bf x}| = O(\sqrt{t})$, for large $t$. It is clearly not valid when $|\vec{x}| = O(1) \ll \sqrt{t}$. In this range, $p(\vec{x},t) \sim O(1)$. For example, exactly at $\vec{x}=\vec{0}$, by definition, $p(\vec{0},t) = 1$. Thus, in the sum $\langle C_N(t) \rangle = \sum_{\vec{x}} \left[p(\vec{x},t)\right]^N$, we can separate the non-scaling part and the scaling part. For the scaling part, we can again use the third line of Eq. (\ref{summary_pxt}) but the integral over $z$ is cut off at $z =a/\sqrt{4\,D\,t}$, where $a$ is a short-distance cut-off. Using the small $z$ behaviour of the integrand $z^{d-1} \, [f_>(z)]^N \sim z^{2N - d(N-1)-1}$ (as discussed above), one finds that the contribution from the lower limit scales for large $t$ as $t^{-\nu}$ where $\nu$ is given in Eq. (\ref{CN_dge2.2}). This exactly cancels the prefactor $\sim t^{\nu}$ in Eq. (\ref{CN_dge2}). Hence for large $t$, we get 
\bea \label{CN_dgeq2}
\langle C_N(t)\rangle \simeq  {\rm const.} \quad {\rm for} \quad d > d_c(N) \;,
\eea
where the constant evidently depends on the cut-off, i.e., on the details of the lattice and hence is non universal. Physically, this indicates that for $d>d_c(N)$, the
common sites visited by all the walkers are typically close to the origin and they get visited at relatively early times. At late times, the walkers hardly overlap and hence $\langle C_N(t) \rangle$ ceases to grow with time. 

\hspace*{0.cm}Exactly at $d =d_c(N)$, the exponent $\nu = 0$ from Eq. (\ref{nu}). Using the asymptotic small $z$ behavior of $f_>(z)$ from the first line of Eq. (\ref{asympt_f>}), one finds that the integrand in Eq. (\ref{CN_dge2}) behaves as $[\Gamma(1/(N-1))]^{1/N}\, z^{-1}$. Note that in this case also, we need to keep the lower limit as $a/\sqrt{4 D t}$ with a cut-off $a$. Consequently, at large time, the integral in Eq. (\ref{CN_dge2}) is dominated by the lower limit $z \to 0$ and we get for large $t$
\be \label{CN_dc}
\langle C_N(t)\rangle \simeq  a_c(N)\, \ln t\;, \quad {\rm with} \quad a_c(N) = \frac{1}{2}  \left[\frac{E_{d_c} \, \Gamma\left( \frac{1}{N-1}\right)}{(4 \pi D)^{d_c/2}}\right]^N (4\,D)^{d_c/2}
\ee 
where we used the shorthand notation $d_c\equiv d_c(N)$. 

\hspace*{0.cm}Summarising the large $t$ behavior of $\langle C_N(t) \rangle$ for $d>2$, we obtain
\bea \label{summary_dge2} 
\langle C_N(t)\rangle \simeq 
\begin{cases}
&\alpha_N(d)\, t^\nu \quad, \quad \hspace*{0.3cm} 2 < d < d_c(N) \;, \\
& \\
& a_c(N) \, \ln t \quad, \quad d = d_c(N) \;, \\
& \\
& {\rm const.} \quad, \quad \hspace*{0.5cm} d > d_c(N) \;,
\end{cases}
\eea
where the expression for the amplitudes $\alpha_N(d)$ and $a_c(N)$ are given above and $\nu =  N- \frac{d(N-1)}{2}$. 
}

\item[(iii)]{\hspace*{0.cm}$d=2$: In this case, we substitute the scaling form of $p(\vec{x},t)$ from the second line of Eq. (\ref{summary_pxt}) into the expression for $\langle C_N(t)\rangle$ in Eq. (\ref{av_CN}), replace the sum by an integral as usual and use the spherical symmetry to obtain
\bea \label{CNdeq2}
\langle C_N(t)\rangle \simeq b_2(N) \; \frac{t}{(\ln t)^N} \quad, \quad b_2(N) = 8 \pi D \int_0^\infty \left[ f_2(z)\right]^N\, z \,dz \;,
\eea
where the function $f_2(z)$ is given in Eq. (\ref{p_eq2}). 

}

\end{itemize}
 
\begin{figure}[t]
\centering
\includegraphics[width = 0.5 \linewidth]{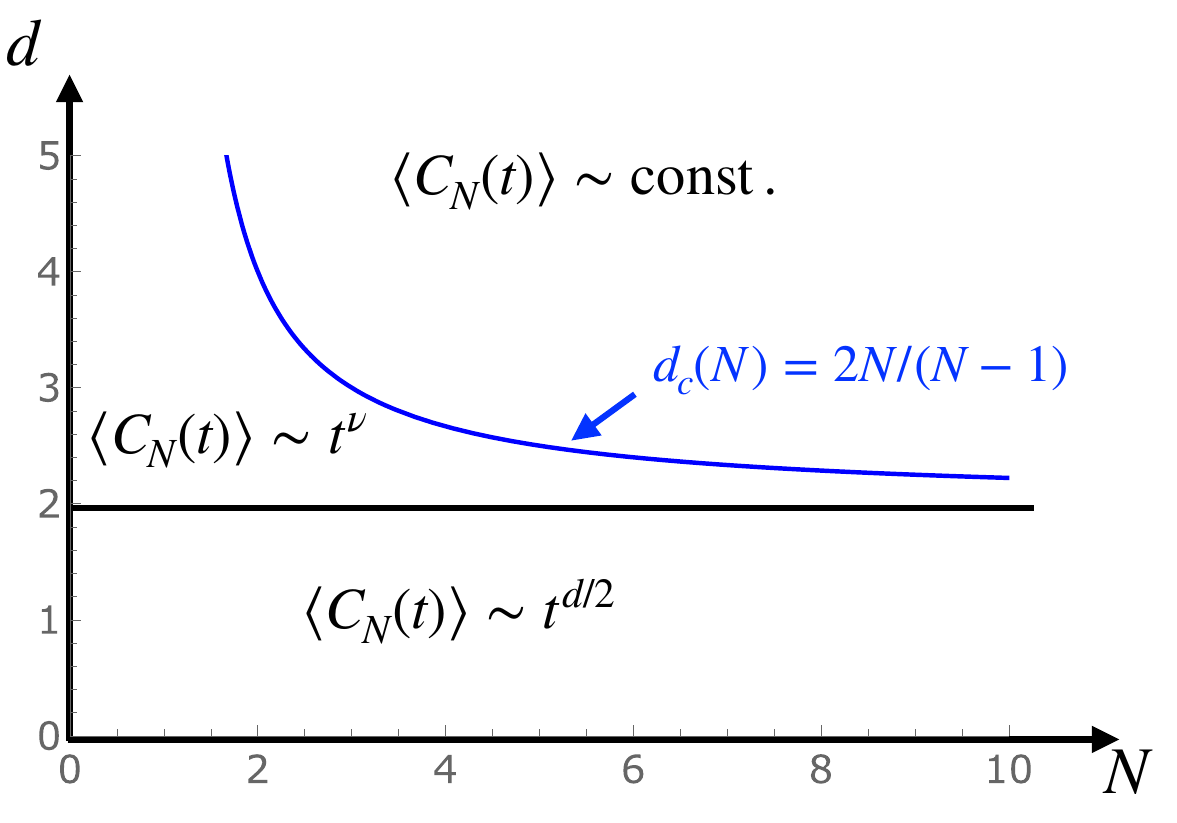}
\caption{Phase diagram in the $(N,d)$ plane for the asymptotic large $t$ growth of the mean number of common sites $\langle C_N\rangle$ visited by
$N$ independent random walkers in $d$ dimensions. There are two critical lines: $d=2$ (the black horizontal line) and $d_c(N) = 2N/(N-1)$ that separate three
regimes of growth. In the regime $2<d<d_c(N)$, one obtains an algebraic growth $\langle C_N(t)\rangle \sim t^{\nu}$ where the exponent $\nu = N-(d/2)(N-1)$ is anomalous and depends on the parameters $d$ and $N$.   
} \label{Fig_phdiag}
\end{figure}

Let us summarize the main results discussed above, that originally appeared in Ref. \cite{Majtam}. 
One obtains a host of rich growth behavior of $\langle C_N(t) \rangle$ for large $t$, depending on the two parameters $d$ and $N$. These growth laws are summarised in the phase diagram in the $(N,d)$ plane in Fig. \ref{Fig_phdiag}. Even though $N$ and $d$ are integers, and so is $C_N(t)$, once we have formulated the problem in terms of $p(\vec{x},t)$ (the probability to visit a site $\vec{x}$ before time $t$), one can consider the mean $\langle C_N(t) \rangle$ as a continuous variable. Moreover it can be analytically continued to real $d$ (for example to fractal lattices) and also to real $N$. One of the most interesting result of this analysis is the existence of this anomalous regime $2<d<d_c(N)$ regime in the $(N,d)$ plane (see Fig.~\ref{Fig_phdiag}) where $\langle C_N(t)\rangle \sim t^{\nu}$ grows algebraically for large $t$, but with an exponent $\nu = N-(d/2)(N-1)$ that depends continuously on $d$ and $N$. For example, for $N=2$, the critical dimension $d_c(2) = 4$. Hence $d=3$ would fall in this anomalous regime where $\langle C_2(t) \rangle t^{1/2}$. This analysis also shows that exactly at $d=d_c(N)$, the mean number $\langle C_N(t) \rangle \sim \ln t$ grows very slowly with time. For example, for $N=3$, where $d_c(3)=3$, if we consider $d=3$, then our results predict $\langle C_3(t) \rangle \sim \ln t$. Some of these predictions have been verified in numerical simulations in Ref. \cite{Majtam}.

 \subsection{Asymptotic behavior of $\langle V_{K,N}(t)\rangle$}

 In this Section, we derive the asymptotic late time growth of $\langle V_{K,N}(t)\rangle$ denoting the mean number of sites visited by exactly $K$ out of the $N$ walkers up to time $t$ (with $1 \leq K \leq N$). The exact formula for $\langle V_{K,N}(t)\rangle$ in terms of the central quantity $p(\vec{x},t)$ is given in Eq. (\ref{VK}), which already appeared in Ref. \cite{Majtam}, but it was not analysed for general~$K$. We recall that for $K=N$, one recovers the mean number of common sites $\langle V_{N,N}(t)\rangle = \langle C_{N}(t)\rangle$ analysed in the previous subsection. In this subsection, we extend this analysis for other values of $K$. Our strategy again is to inject the asymptotic scaling behavior of $p(\vec{x},t)$ from Eq. (\ref{summary_pxt}) into Eq. (\ref{VK}) valid for large $t$ and then analyse the sum in Eq. (\ref{VK}) upon replacing it by an integral and computing it using spherical symmetry.  The details are exactly similar as in the case $K=N$ case and therefore we provide only the main results. As in the $K=N$ case, it turns out that $d=2$ is a critical line and we consider the three regimes separately: (i) $d<2$, (ii) $d>2$ and (iii) $d=2$. 
 
 \begin{itemize}
 
 \item[(i)]{$d<2$: We start with the simplest case $d<2$. Following the strategy mentioned above, it is easy to verify that, for large $t$, 
 \bea \label{VK_dless2}
 \langle V_{K,N}(t) \rangle \sim t^{d/2} \quad {\rm for \; all} \; 1 \leq K \leq N \;.
 \eea
 Thus the growth exponent $d/2$ is independent of $K$, while the $K$ and $N$-dependence appear only in the amplitude, which can in principle be computed. 
}

\item[(ii)]{\hspace*{0.cm}$d>2$: In this case, $p(\vec{x},t) \sim t^{1-d/2} f_>(|\vec{x}|/\sqrt{4 Dt})$ from the third line of Eq. (\ref{summary_pxt}). Substituting this behaviour in Eq. (\ref{VK}), one finds that, due to the decaying prefactor $t^{1-d/2}$ of $f_>(|\vec{x}|/\sqrt{4 Dt})$, at large times, one can approximate 
$(1 - p(\vec{x},t))^{N-k} \simeq 1$. Consequently, Eq. (\ref{VK}) reduces to
\bea \label{VKN_ge2}
\langle V_{K,N}(t) \rangle \simeq {N \choose K} \sum_{\vec{x}} \left[p(\vec{x},t) \right]^K \;.
\eea 
Thus, up to this prefactor ${N \choose K}$, this is exactly identical to Eq. (\ref{av_CN}) with $N$ replaced by $K$. Thus, all the scaling analysis done before for $\langle C_N(t) \rangle$ will go through after the replacement of $N$ by $K$. In particular we thus get a critical line $d_c(K) = 2K/(K-1)$. Thus replacing $N$ by $K$ in Eq. (\ref{summary_dge2}), we get for $K>1$, up to unimportant prefactors
 \bea \label{summary_VKN_dge2} 
\langle V_{K,N}(t)\rangle \sim 
\begin{cases}
& t^\nu \quad, \quad \hspace*{1.3cm} 2 < d < d_c(K) \;, \\
& \\
&  \ln t \quad, \quad \hspace*{1.cm} d = d_c(K) \;, \\
& \\
& {\rm const.} \quad, \quad \hspace*{0.5cm} d > d_c(K) \;,
\end{cases}
\eea
where $\nu = K - (d/2)(K-1)$. The case $K=1$ is special because for this case $d_c(1) \to \infty$ and $\nu = 1$. Indeed, in this case, from Eq. (\ref{VKN_ge2}), one finds 
\bea \label{V1}
\langle V_{1,N}(t) \rangle \simeq N \sum_{\vec x} p(\vec{x},t) = N \langle D_1(t) \rangle\;,
\eea
where the last equality follows from Eq. (\ref{av_D1}). Note that $D_1(t)$ is just the number of distinct sites visited by a single walker. Using Eqs. (\ref{DNge2.4}) and (\ref{DNge2.5}) for $N=1$, one finds that $\langle D_1(t) \rangle \simeq E_d\, t$ for large $t$ where $E_d$ is the escape probability of a single walker. Consequently, one gets, from Eq. (\ref{V1})
\bea \label{V1.1}
\langle V_{1,N}(t) \rangle \simeq N  \,E_d\, t \;,
\eea
for large $t$. This result implies that for $d >2$, each site is visited on an average by only one walker and only once. 

}

\item[(iii)]{\hspace*{0.cm}$d=2$: In this case, using $p(\vec{x},t) \sim [1/\ln(t)]\, f_2\left(|\vec{x}|/\sqrt{4 D t}\right)$ and approximating $(1 - p(\vec{x},t))^{N-k} \simeq 1$, one finds, after replacing $N$ by $K$ in Eq. (\ref{CNdeq2}), that at late times
\bea \label{V2_deq2}
\langle V_{K,N}(t)\rangle \sim \frac{t}{(\ln t)^K} \;.
\eea

}
 
 \end{itemize}
 
Thus, for each $1<K\leq N$, one can draw a phase diagram in the $(K,d)$ plane, similar to the $K=N$ case in Fig.~\ref{Fig_phdiag}. Essentially, there are three regimes in the $(K,d)$ plane, where the asymptotic behaviors are given by
\bea
\langle V_{K,N}\rangle
\sim
\begin{cases}
&t^{d/2} \quad, \quad d<2\:, \\
& \\
& t^{\nu} \quad, \quad 2<d<d_c(K)\;, \\
& \\
& {\rm const.} \quad, \quad d>d_c(K) \;,
\end{cases}
\eea 
where $\nu = K - (d/2)(K-1)$ and $d_c(K) = 2K/(K-1)$. Exactly at $d=d_c(K)$, the mean value grows as $\langle V_{K,N}\rangle \sim \ln t$, while at $d=2$ it behaves as $\sim t/(\ln t)^K$ for large $t$.

\section{The exact distributions of $D_N(t)$ and $C_N(t)$ in one dimension and the link to extreme value statistics}

\begin{figure}[t]
\centering
\includegraphics[width = 0.7\linewidth]{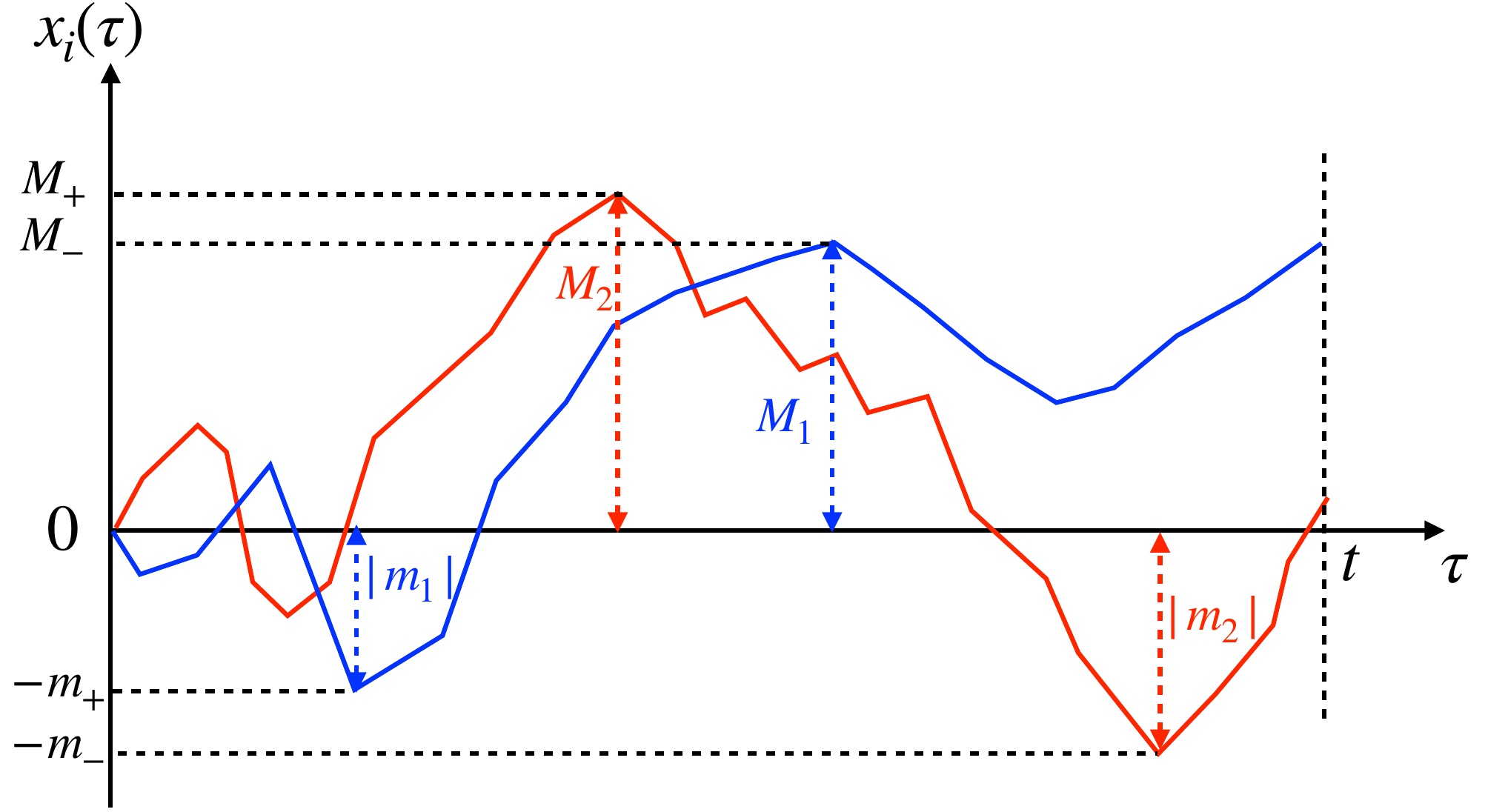}
\caption{Schematic trajectories of $N=2$ independent Brownian motions, each of duration $t$. The The variables $M_1$, $M_2$ (denoting the global maximum of each walker) and $m_1, m_2$ (denoting the global minimum of each walker) are indicated. Similarly, $M_+ = \max(M_1, M_2)$, $M_- = \min(M_1, M_2)$, $m_- = \max(|m_1|, |m_2|)$ and $m_+ = \min(|m_1|, |m_2))$ are also indicated.}\label{fig_1d}
\end{figure}

In the previous sections, we have calculated exactly the mean number of distinct and common sites visited up to time $t$ by $N$ independent
random walks in $d$ dimensions. For large time, these random walks converge to Brownian motions and these two observables become easier to compute
in the Brownian limit. However, calculating the higher moments or the full distributions of $D_N(t)$ (the number of distinct sites up to $t$)
and $C_N(t)$ (the number of common sites up to $t$) is a very difficult problem in arbitrary dimension. However, in $d=1$, one can compute the 
full distribution of $D_N(t)$ and $C_N(t)$~\cite{KMS2013}. In this section, we briefly outline the salient features leading to these exact results.

We consider $N$ independent Brownian motions each of duration $t$ and each starting at the origin. Let $x_i(\tau)$ denote the position of the $i$-th walker at time $\tau$ (with $0\leq \tau \leq t$) -- see Fig. \ref{fig_1d} for $N=2$. For each walk, we first identify the global maximum $M_i$ and the global minimum $m_i$. Note that the maxima $M_i$'s are necessarily non-negative, i.e., $M_i \geq 0$, while the minima $m_i$'s are necessarily non-positive, i.e., $m_i \leq 0$. Let us now define the pair of variables $M_+$ and $m_-$ 
\bea \label{defM+m-}
M_+ &=& \max\{M_1, M_2, \cdots, M_N \} = \max_{1 \leq i \leq N} M_i  \;, \\
m_- &=& -\min \{m_1, m_2, \cdots, m_N \} = \max\{|m_1|, |m_2|, \cdots, |m_N| \} = \max_{1 \leq i \leq N} |m_i|  \,.\nonumber \\
\eea
Here $M_+$ denotes the global maximum of all the $N$ Brownian motions, while $m_-$ denotes the absolute value of the global minimum of these walks up to time $t$. For brevity of notations, we do not exhibit the explicit time-dependence of these random variables. The number of distinct sites visited up to time $t$ by all the walkers on the positive side of the origin, in this continuous time limit, is just $M_+(t)$, since it represents the span of the walkers on the positive side. Similarly $m_-(t)$ represents the number of distinct sites visited by the walkers on the negative side up to time $t$. Hence, the number of distinct sites $D_N(t)$ is just the sum of these two observables
\bea \label{DN_M+m-}
D_N(t) = M_+ + m_- \;. 
\eea
Similarly, let us define the pair of variables $M_+$ and $m_-$ 
\bea \label{defM-m+}
M_- &=& \min\{M_1, M_2, \cdots, M_N \} = \min_{1 \leq i \leq N} M_i  \;, \\
m_+&=& -\max \{m_1, m_2, \cdots, m_N \} = \min\{|m_1|, |m_2|, \cdots, |m_N| \} = \min_{1 \leq i \leq N} |m_i|  \,. \nonumber \\
\eea
These four observables $M_+, M_-, m_+$ and $m_-$ are indicated in Fig. \ref{fig_1d} for $N=2$ walkers.

The variable $M_-$ denotes the smallest of the $N$ maxima, while $m_+$ denotes the smallest of the absolute values of the $N$ minima. Therefore, $M_-$ denotes the intersection of all the sites visited by $N$ walkers on the positive side of the origin, while $m_+$ denotes the intersection on the negative side. Hence, the number of common sites $C_N(t)$ visited by all the $N$ walkers up to time $t$ is given by the sum
\bea \label{CN_M-m+}
C_N(t) = M_- + m _+ \;.
\eea

Thus, in order to compute the distribution of $D_N(t)$ using the relation in Eq. (\ref{DN_M+m-}), we need the joint distribution of $M_+$ and $m_-$. Similarly, for the distribution of $C_N(t)$ in Eq. (\ref{CN_M-m+}), we need the joint distribution of $M_-$ and $m_+$. These joint distributions can be computed explicitly using the fact that the walkers are independent. This brings us to the extreme value statistics (EVS) of $N$ independent and identically distributed (IID) random variables \cite{us_book,EVS_review}. To proceed, we first make a simple observation that, for a single Brownian motion, the joint distribution of $M_i$
 and $m_i$ is only a function of the rescaled variables $\widetilde{M}_i = M_i/\sqrt{4\,D\,t}$ and $\widetilde{m}_i=m_i/\sqrt{4\,D\,t}$ \cite{EVS_review}. In other words, the dependence on $t$ only appears through these rescaled variables. For simplicity, we set $D=1/2$ and define the rescaled variables 
 \bea \label{def_rescaled}
 \widetilde{M}_{\pm} = \frac{M_{\pm}}{\sqrt{2t}} \quad, \quad \widetilde{m}_{\pm} = \frac{m_{\pm}}{\sqrt{2t}}  \;.
 \eea
 In terms of the rescaled variables, one can think of $N$ Brownian motions defined on the unit time interval. In the following two subsections, we compute the distribution of $D_N(t)$ and $C_N(t)$ separately.

 \subsection{Distribution of the number of distinct sites} \label{sec:PDFdist}

 Let us start with the computation of the PDF of $D_N(t)$ in Eq. (\ref{DN_M+m-}). For this, we need 
 to compute the joint distribution of $M_+$ and $m_-$. It turns out to be convenient to consider their cumulative distribution which, in terms of the 
rescaled variables, is defined as
\bea \label{joint_CDFD}
P_D(\ell_1, \ell_2) = {\rm Prob.}\left(\widetilde{M}_{+} \leq \ell_1, \widetilde{m}_{-} \leq \ell_2 \right) \;.
\eea  
If we know this joint cumulative distribution, the joint PDF is simply given~by
\bea \label{joint_PDFD}
p_D(\ell_1, \ell_2) = \frac{\partial^2 }{\partial \ell_1 \partial \ell_2} P_D(\ell_1, \ell_2) \;.
\eea
Using the independence of the $N$ Brownian motions, the cumulative distribution is given~by
\bea \label{CDF_explD}
P_D(\ell_1, \ell_2) = \left[g(\ell_1, \ell_2)\right]^N \;,
\eea 
where  
\bea \label{def_g}
g(\ell_1, \ell_2) = {\rm Prob.}(\widetilde{M} \leq \ell_1, \widetilde{m} \geq - \ell_2) \;.
\eea
Here $\widetilde{M}$ and $\widetilde{m}$ denote respectively the maximum and the minimum of a single Brownian motion on the unit time interval. This joint cumulative distribution for a single Brownian motion can be computed explicitly by solving the diffusion equation in a box $[-\ell_2, \ell_1]$ with absorbing boundary conditions at the two edges \cite{Rednerbook,us_book}. It reads
\bea \label{g_expl}
g(\ell_1, \ell_2) = \frac{2}{\pi} \sum_{n=0}^\infty \frac{1}{n + \frac{1}{2}} \sin\left(\frac{2(n+1)\pi \ell_2}{\ell_1 + \ell_2} \right) \exp{\left[- \left(\frac{(n+\frac{1}{2})\pi}{\ell_1 + \ell_2}\right)^2 \right]} \;.
\eea
Therefore using Eqs. (\ref{DN_M+m-}), (\ref{joint_PDFD}) and (\ref{CDF_explD}), the PDF of $D_N(t)$ is given by
\be \label{pND}
p_N^D(s) = {\rm Prob.} (D_N(t) = s) = \int_0^\infty d \ell_1   \int_0^\infty d \ell_2 \, \delta(s - \ell_1 - \ell_2) \frac{\partial^2}{\partial \ell_1 \partial \ell_2} \left[g(\ell_1, \ell_2)\right]^N.
\ee
For small of values of $N$, one can compute this double integral explicitly and numerical simulations confirm this result~\cite{KMS2013}. For general $N$, it is hard to compute the distribution explicitly from Eq. (\ref{pND}). However, the tails of the distribution for small and large $s$ can be computed for general $N$ and they are given by (details can be found in Ref.~\cite{KMS2013} cite Erratum). For $N \geq 2$, one gets
\bea  \label{asympt_pND}
p_N^D(s) \simeq
\begin{cases}
&a_N \,s^{-5} \, \exp{\left[- N \pi^2/(4\,s^2) \right]} \quad, \quad s \to 0 \\
& \\
&b_N \, \exp{(-s^2/2)} \quad, \quad \hspace*{1.6cm}s \to \infty \;,
\end{cases}
\eea
where the prefactors $a_N$ and $b_N$ are given by
\bea \label{aN_bN}
a_N = 4 \pi^{3/2} N(N-1) \left(\frac{4}{\pi}\right)^{N-2} \frac{\Gamma\left( \frac{N-1}{2}\right)}{\Gamma \left(\frac{N}{2} \right)} \quad, \quad b_N = \frac{2^{3/2}}{\sqrt{\pi}}N(N-1) \;.
\eea
For $N=1$ the asymptotic behaviors are the same as in Eq. (\ref{asympt_pND}) with the prefactors $a_1 = 2 \pi^2$ and $b_1 = 8/\sqrt{\pi}$.

It turns out that, for large $N$, an interesting scaling behavior emerges. 
The typical scale of the  fluctuations of $D_N(t)/\sqrt{2 t}$ can be estimated from the connection to the EVS of IID variables, using Eqs. (\ref{DN_M+m-}) and (\ref{defM-m+}). The rescaled variables $\widetilde{M}_i$'s which are the maxima of the $i^{\rm th}$ 
Brownian motion on the unit interval, are IID variables. Their common PDF is a half-Gaussian \cite{us_book}, $p(M)=({2}/{\sqrt{\pi}}) e^{-M^2}, M>0$.
The same holds for the rescaled variables $-\widetilde{m}_i$'s. Hence, for large $N$, standard results of EVS \cite{EVS_review,Gumbel} state that the typical value of 
$\widetilde M_+ =\max_{1\leq i \leq N} \widetilde M_i$ is   
${O}(\sqrt{\log N})$ while its fluctuations are of order $1/\sqrt{\log N}$ and governed by a Gumbel distribution. This means that, to leading order for large $N$, the random variable $\widetilde{M}_+$ can be expressed as
\bea \label{Gumbel.1}
\widetilde{M}_+ \simeq \sqrt{\ln N} + \frac{1}{2 \sqrt{\ln N}} G_1 \;,
\eea
where $G_1$ is a random variable of $O(1)$ which is distributed via the Gumbel law 
\bea \label{CDF_Gumbel}
{\rm Prob.}(G_1 \leq \ell_1) = e^{-e^{-\ell_1}} \;.
\eea
\begin{figure}[t]
\centering
\includegraphics[width = 0.5\linewidth]{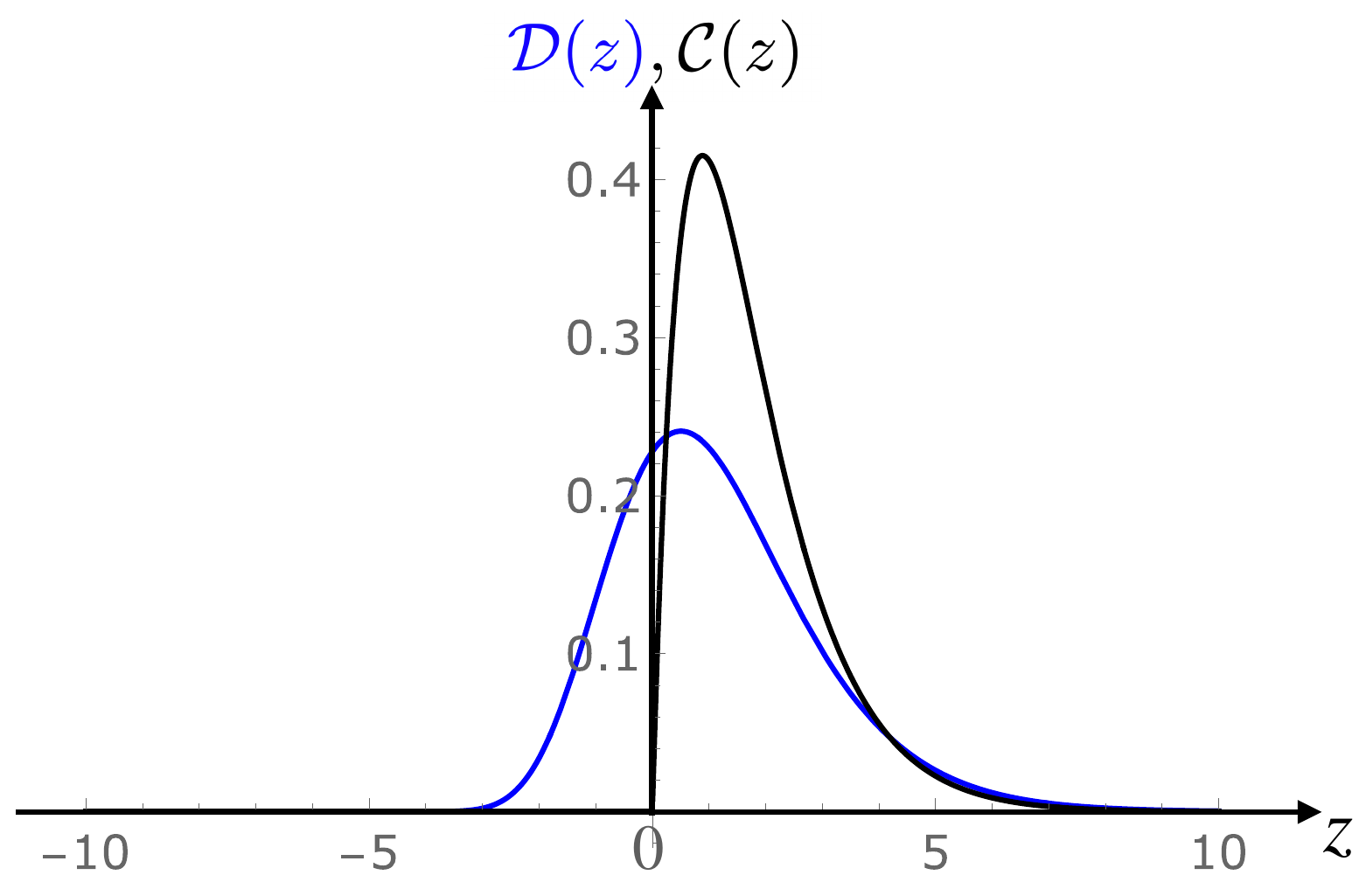}
\caption{Plot of the scaling function ${\cal D}(z)$ -- blue solid line -- given in Eq. (\ref{eq:Bessel2}) and of ${\cal C}(z)$ -- black solid line -- given in Eq. (\ref{pC_scaling}) as a function of $z$.}\label{fig_D}
\end{figure}

Similarly, one can express 
\bea \label{Gumbel.2}
\widetilde{m}_- \simeq \sqrt{\ln N} + \frac{1}{2 \sqrt{\ln N}} G_2 \;,
\eea
where $G_2$ is again distributed via the Gumbel law. For large $N$, these two extremes $\widetilde{M}_+$ and $\widetilde{m}_- $
become uncorrelated as the global maximum and global minimum are most likely not reached by the same walker. Hence, to leading order for large $N$, the variables $\widetilde M_+$ and $\widetilde m_-$ can be considered as independent. This implies that the associated Gumbel variables $G_1$ and $G_2$ are also uncorrelated. Consequently we can write 
\begin{equation}
g^N\left[\mu_N+\tfrac{\tilde{\ell}_1}{2 \mu_N},\mu_N+\tfrac{\tilde{\ell}_2}{2\mu_N}\right] 
\underset{N\to +\infty}{\longrightarrow} e^{-e^{-\widetilde{\ell}_1}}e^{-e^{-\tilde{\ell}_2}}\label{limit_gN}
\end{equation}
with $\mu_N = \sqrt{\ln N}$. Inserting (\ref{limit_gN}) in (\ref{pND}) with $\widetilde{s}=2\mu_N(s-2\mu_N)$ one finds
\begin{equation}
p_N^D(s) \sim 2\sqrt{\ln N}\int_{-\infty}^{\infty} d\widetilde{\ell}_2~e^{-\widetilde{s}}e^{-e^{-\widetilde{\ell}_2}}
e^{-e^{-(\widetilde{s}-\widetilde{l}_2)}} \;, \label{limit-dist-dist}
\end{equation}
which can be evaluated explicitly. This gives the scaling form for the PDF of $D_N(t)$ for large $N$ 
\bea \label{eq:Bessel}
p_N^D(s) \simeq 2 \sqrt{\ln N}\, {\cal D}(2 \sqrt{\ln N}(s-2\ln N)) \;,
\eea
where the scaling function ${\cal D}(z)$ is given by
\bea \label{eq:Bessel2}
{\cal D}(z) = 2\, e^{-z}\, K_0(2\,e^{-z/2}) \;.
\eea
Here $K_0(z)$ is the modified Bessel function of index $0$. This result was verified in numerical simulations for $N=50$ and $N=100$ in Ref. \cite{KMS2013}. Here we just provide a plot of the scaling function ${\cal D}(z)$ in Fig. \ref{fig_D}. Its asymptotic behaviors are given by
\bea \label{asympt_D}
{\cal D}(z) \simeq
\begin{cases}
&\sqrt{\pi} \, e^{{3|z|/4}-2 e^{|z|/2}} \quad, \quad z \to - \infty \\
& \\
& z\,e^{-z} \quad, \hspace*{1.7cm}\quad z \to \infty \;.
\end{cases}
\eea 
This fact that the convolution of two independent Gumbel variables is distributed via the modified Bessel function also appeared in other contexts, e.g., for the maximum of a log-correlated gas of particles on a circle \cite{FB08,SZ2015}.

\subsection{Distribution of the number of common sites} \label{sec:PDFcom}

We now turn to the PDF of $C_N(t)$ in Eq. (\ref{CN_M-m+}). This requires the computation of the joint PDF of $M_-$ and $m_+$. 
It turns out to be convenient again to consider the cumulative distribution of the rescaled variables 
\bea \label{joint_CDFC}
P_C(j_1, j_2) = {\rm Prob.}\left(\widetilde{M}_{-} \geq j_1, \widetilde{m}_{+} \geq j_2 \right) \;.
\eea  
If we know this joint cumulative distribution, the joint PDF is simply given~by
\bea \label{joint_PDFC}
p_C(j_1, j_2) = \frac{\partial^2 }{\partial j_1 \partial j_2} P_C(j_1, j_2) \;.
\eea
Using the independence of the $N$ Brownian motions, the cumulative distribution is given by
\bea \label{CDF_explC}
P_C(j_1, j_2) = \left[h(j_1, j_2)\right]^N \;,
\eea 
where  
\bea \label{def_h}
h(j_1, j_2) = {\rm Prob.}(\widetilde{M} \geq j_1, |\widetilde{m}| \geq j_2) =  {\rm Prob.}(\widetilde{M} \geq j_1, \widetilde{m} \leq - j_2)\;,
\eea
where $\widetilde{M}$ and $\widetilde{m}$ denote respectively the maximum and the minimum of a single Brownian motion on the unit time interval. In writing the last equality in Eq. (\ref{def_h}) we used the fact that $\tilde m \leq 0$. In fact, by using inclusion--exclusion principle of probability, it is easy to see that  
$h(j_1, j_2)$ is related to the function $g(\ell_1, \ell_2)$ in Eq. (\ref{g_expl}) via 
\bea \label{rel_gh}
h(j_1, j_2) =  1- {\rm erf}(j_1) - {\rm erf}(j_2) + g(j_1,j_2) \;,
\eea 
where ${\rm erf}(x) = 2/\sqrt{\pi}\int_0^x e^{-u^2\,du}$. Note that, here, we used the fact that ${\rm Prob.}(\tilde M \leq j_1)={\rm erf}(j_1)$ and ${\rm Prob.}(\tilde m\geq - j_2) = {\rm erf}(j_2)$ \cite{EVS_review,us_book}. Therefore using Eqs. (\ref{CN_M-m+}), (\ref{joint_PDFC}) and (\ref{CDF_explC}), the PDF of $C_N(t)$ is given by
\bea \label{pNC}
p_N^C(w) = {\rm Prob.} (C_N(t) = w) = \int_0^\infty d \ell_1   \int_0^\infty d \ell_2 \, \delta(w - j_1 - j_2) \frac{\partial^2}{\partial j_1 \partial j_2} \left[h(j_1, j_2)\right]^N \;.
\eea
The asymptotic behavior of $p^C_N(w)$ for $N \geq 2$ are given by \cite{KMS2013}
\bea \label{asympt_pNC}
p_N^C(w) \simeq
\begin{cases}
& c_N \, w \quad, \quad \hspace*{2.6cm} w \to 0 \\
& \\
& d_N \,w^{1-N} \exp{(-N w^2}) \quad, \quad w \to \infty \;,
\end{cases}
\eea
where the constants $c_N$ and $d_N$ are given by
\bea \label{cN_dN}
c_N = \frac{4}{\pi}N(N-1) \quad, \quad d_N = \frac{8N}{\pi^{N/2}} \;.
\eea
For $N=1$, i.e., for a single Brownian motion, the number of distinct and common sites are identical, i.e., $D_1(t) = C_1(t)$. Hence the asymptotic behavior of $p_1^C(w)$ can be read off Eq. (\ref{asympt_pND}) with $N=1$, for which $a_1 = 2 \pi^2$ and $b_1 = 8/\sqrt{\pi}$. 

As in the case of $D_N(t)$, the PDF of $C_N(t)$ also exhibits an interesting scaling behavior for large $N$. The
The typical scale of the  fluctuations of $C_N(t)/\sqrt{2 t}$ can be estimated from the connection to the EVS of IID variables, using Eqs. (\ref{CN_M-m+}) and (\ref{defM+m-}). The rescaled variables $\widetilde{M}_i$'s which are the maxima of the $i^{\rm th}$ 
Brownian motion on the unit interval, are IID variables. Their common PDF is a half-Gaussian \cite{us_book}, $p(M)=({2}/{\sqrt{\pi}}) e^{-M^2}, M>0$.
The same holds for the rescaled variables $-\widetilde{m}_i$'s. Hence, for large $N$, standard results of EVS \cite{Gumbel,EVS_review} state that the PDF of their minimum scales as $\widetilde{M}_- \simeq X/N$ where the random variable $X$ is distributed via the Weibull law
\bea \label{Weibull}
{\rm Prob.}(X \geq z) = e^{-z/\sqrt{\pi}} \;.
\eea 
Similarly, for large $N$, we have $m_+ \simeq Y/N$ where $Y$ is also distributed by the same law as in Eq. (\ref{Weibull}). For large $N$, the minimum on the positive side and the minimum on the negative side are not achieved by the same walker. Hence $X$ and $Y$ can be considered as independent random variables, each of which being distributed via (\ref{Weibull}). Consequently, their sum $C_N(t)$ is a convolution of two exponentials and it is easy to see that the PDF of $C_N(t)$ can be written in the scaling form
\bea \label{pC_scaling}
p_N^C(w) \simeq N\; {\cal C}(N\, w)
\eea
where the scaling function ${\cal C}(z)$ is given by
\bea \label{lim_exp}
{\cal C}(z) = \frac{4}{\pi}\,z \,e^{- 2 z/\sqrt{\pi}} \;, \quad z \geq 0 \;.
\eea
This scaling function is plotted in Fig. \ref{fig_D}, along with ${\cal D}(z)$.

\section{Extension to other models}

The method presented here for computing the statistics of $D_N(t)$ and $C_N(t)$ for $N$ independent Brownian walkers can be easily extended to other
models where the walkers remain independent but their individual stochastic motion need not be Brownian. We briefly discuss two simple examples below. 

\subsection{$N$ independent Brownian bridges}

A Brownian bridge is a Brownian motion which is constrained to come back to its starting point (here the origin) after a fixed time $t$. 
In models of animals foraging, the Brownian bridge often plays an important role since animals typically come back to their nest at the
end of the day, after foraging. It is then natural to ask what is the number of distinct and common sites visited by $N$ Brownian bridges
in $d$ dimensions. The method presented in Section~\ref{sec:method} for $N$ Brownian motions involve the central quantity $p(\vec{x},t)$
denoting the probability that the site $\vec{x}$ is visited by a single walker before time $t$. The same method, in terms of $p(\vec{x},t)$, goes through
for $N$ independent Brownian bridges except that $p(\vec{x},t)$ for a Brownian bridge of duration $t$ is not exactly identical to that of a Brownian motion.
However, one can show that $p(\vec{x},t)$ for large $|\vec{x}|$ and large $t$, for a Brownian bridge, exhibits exactly similar scaling forms as in Eq. (\ref{summary_pxt}) for the Brownian motion, except that the scaling functions $f_<(z)$ (for $d<2$), $f_2(z)$ (for $d=2$) and $f_>(z)$ (for $d>2$) for the Brownian bridges are different from their Brownian motion counterparts. Therefore the large time behavior of $\langle D_N(t) \rangle$ and $\langle C_N(t) \rangle$ for $N$ Brownian bridges will be exactly similar to those of the Brownian motions, respectively in Eqs. (\ref{DN_summary}) and (\ref{summary_dge2}), except that the prefactors will be different. Consequently the phase diagram shown in Fig. \ref{Fig_phdiag} will also be similar for the Brownian bridges. 

Furthermore, in $d=1$, one can compute the distribution of $D_N(t)$ and $C_N(t)$ for $N$ independent Brownian bridges, as in the Brownian motion case. For finite $N$ these distributions for the bridges are different from those of Brownian motions. However, for large $N$, using the universality of the EVS of IID random variables \cite{us_book,EVS_review}, it is possible to show that the distribution of $D_N(t)$ and $C_N(t)$ -- up to some trivial scale factors -- are exactly the same as those of Brownian motions, given respectively in Eqs. (\ref{eq:Bessel}) and (\ref{pC_scaling}). Thus the two scaling functions ${\cal D}(z)$ and ${\cal C}(z)$ in Eqs. (\ref{eq:Bessel2}) and (\ref{lim_exp}) are universal, i.e., they are independent of the bridge constraint.

\subsection{$N$ independent resetting Brownian motions}

A resetting Brownian motion (RBM) is a Brownian motion that resets to the origin with a constant rate $r$ and has been studied extensively during the last decade (for a review see \cite{EMS2020}). In the context of animal foraging, RBM can be used to model the fact that an animal can return to the nest from time to time. In this context one can also study the resetting Brownian bridge (RBB) \cite{BMS2022}. For a single RBM, the mean number of distinct sites $\langle D_1(t) \rangle$ has been studied in all dimensions and has been shown to grow extremely slowly as  $\langle D_1(t) \rangle \sim (\ln t)^d$ for large $t$ \cite{BMM2022}. This is considerably slower compared to the Brownian motion and in addition, the special role played by $d=2$ disappears in the case of RBM. In addition, the distribution of $D_1(t)$ has also been computed exactly in $d=1$~\cite{BMM2022}. However, the statistics of $D_N(t)$ and $C_N(t)$ for $N$ independent RBM or RBB has not yet been studied and thus remain an interesting open problem. The method used in this paper may be useful to solve this problem.

\section{Conclusion}

The motion of a foraging animal can often be modelled by a random walker/Brownian motion. In this chapter, we considered a simple model of  
$N$ independent foraging animals, each of them performing independent Brownian motion. Even though each walker is independent, the statistical
properties of the territory covered (for example the home range \cite{GK2014}) by the animals can have nontrivial statistics. There are several measures for characterizing the area of the territory
by the $N$ animals. For example, one popular measure is the convex hull of the union of the trajectories of these Brownian motions and the statistics of this convex hull has been studied extensively \cite{RMC2009,MCR2010,CHM2015,HMSS2020,MMSS2021,SKMS2022}. Another interesting measure of the home range is the number of distinct sites visited by $N$ walkers up to time $t$ \cite{Larralde}. A related question concerns the statistics of the number of sites visited by all the $N$ walkers \cite{Majtam}. In this Chapter we reviewed the recent results on the last two quantities for $N$ independent Brownian motions and also discussed the extension of these results to two other models, namely the $N$ independent Brownian bridges and $N$ independent resetting Brownian motions.   

Let us summarize the main results presented here. We have studied, in all dimensions $d$, the mean number of distinct ($\langle D_N(t) \rangle$) and common sites ($\langle C_N(t) \rangle$) up to time by $N$ independent random walks which, in the large time limit, can be studied in terms of $N$ independent Brownian motions. We have shown that, to compute these two quantities, one needs to study the late time scaling behavior of a central quantity $p(\vec{x},t)$, denoting the probability that the site $\vec{x}$ has been visited by a single walker before time $t$. 
We have shown that $\langle D_N(t)\rangle \sim t^{d/2}$ for $d<2$ and $\langle D_N(t)\rangle \sim t$ for $d>2$, as for a single walker, with $N$-dependent prefactors. In contrast, $\langle C_N(t) \rangle$ exhibits a much richer behavior. While for $d<2$ it grows as $\langle C_N(t) \rangle \sim t^{d/2}$ with an $N$-dependent prefactor, it has very different behaviors for $d>2$. In particular there exists a critical dimension $d_c(N) = 2N/(N-1)$ such that, for $2<d<d_c(N)$, the asymptotic growth is given by $\langle C_N(t) \rangle \sim t^{\nu}$ where the exponent $\nu = N - d(N-1)/2$ depends on $d$ and $N$. For $d>d_c(N)$, the mean number of common sites approaches a constant at late times, while exactly at $d=d_c(N)$ it grows logarithmically with an $N$-dependent prefactor. These behaviors in the $(N,d)$ plane are summarised in the phase diagram in Fig. \ref{Fig_phdiag}. In addition, we have also computed the mean number of sites visited by exactly $1\leq K \leq N$ walkers up to time $t$. The critical dimension in this case is $d_c(K) = 2K/(K-1)$. Finally, in $d=1$, we have computed the full distribution of $D_N(t)$ and $C_N(t)$ for any $N$ and showed that they exhibit an interesting scaling behavior for large $N$. This provides a nice and interesting application of the extreme value statistics. 

The method presented here is general enough to be extended to other types of stochastic processes. For example, we have discussed two applications, 
namely the case of $N$ independent Brownian bridges and $N$ independent resetting Brownian motions/bridges. In particular, the latter case has been studied only for $N=1$ and possible extension to general $N$ is an interesting open problem. Furthermore, the full distributions of $D_N(t)$ and $C_N(t)$ are known only in $d=1$ and computing them for $d>1$ remains a challenging problem. Finally, we considered here $N$ noninteracting Brownian motions. However, in reality the animals are always interacting and it would be interesting to study the statistics of $D_N(t)$ and $C_N(t)$ for interacting random walkers.

\vspace*{0.5cm}   
\noindent{\it Acknowledgments:} We would like to thank M. Biroli, A. Kundu, F. Mori and M. Tamm for useful discussions and collaborations on this subject. We acknowledge support from ANR, Grant No. ANR- 23- CE30-0020-01 EDIPS.

\end{document}